\documentclass[prd,aps,amsmath,amssymb,twocolumn]{revtex4}
\usepackage{graphicx}% Include figure files
\usepackage{dcolumn}% Align table columns on decimal point
\usepackage{bm}% bold math

\begin{document}

%%%%
%    Greek Letters
%

\let\a=\alpha      \let\b=\beta       \let\c=\chi        \let\d=\delta
\let\e=\varepsilon \let\f=\varphi     \let\g=\gamma      \let\h=\eta
\let\k=\kappa      \let\l=\lambda     \let\m=\mu
\let\o=\omega      \let\r=\varrho     \let\s=\sigma
\let\t=\tau        \let\th=\vartheta  \let\y=\upsilon    \let\x=\xi
\let\z=\zeta       \let\io=\iota      \let\vp=\varpi     \let\ro=\rho
\let\ph=\phi       \let\ep=\epsilon   \let\te=\theta
\let\n=\nu
\let\D=\Delta   \let\F=\Phi    \let\G=\Gamma  \let\L=\Lambda
\let\O=\Omega   \let\P=\Pi     \let\Ps=\Psi   \let\Si=\Sigma
\let\Th=\Theta  \let\X=\Xi     \let\Y=\Upsilon

%
%%%

%%%
%    Calligraphic letters
%

\def\cA{{\cal A}}                \def\cB{{\cal B}}
\def\cC{{\cal C}}                \def\cD{{\cal D}}
\def\cE{{\cal E}}                \def\cF{{\cal F}}
\def\cG{{\cal G}}                \def\cH{{\cal H}}
\def\cI{{\cal I}}                \def\cJ{{\cal J}}
\def\cK{{\cal K}}                \def\cL{{\cal L}}
\def\cM{{\cal M}}                \def\cN{{\cal N}}
\def\cO{{\cal O}}                \def\cP{{\cal P}}
\def\cQ{{\cal Q}}                \def\cR{{\cal R}}
\def\cS{{\cal S}}                \def\cT{{\cal T}}
\def\cU{{\cal U}}                \def\cV{{\cal V}}
\def\cW{{\cal W}}                \def\cX{{\cal X}}
\def\cY{{\cal Y}}                \def\cZ{{\cal Z}}

\def\dbd{{$0\nu 2\beta\,$}}
%
%%%%

\newcommand{\Ns}{N\hspace{-4.7mm}\not\hspace{2.7mm}}
\newcommand{\qs}{q\hspace{-3.7mm}\not\hspace{3.4mm}}
\newcommand{\ps}{p\hspace{-3.3mm}\not\hspace{1.2mm}}
\newcommand{\ks}{k\hspace{-3.3mm}\not\hspace{1.2mm}}
\newcommand{\des}{\partial\hspace{-4.mm}\not\hspace{2.5mm}}
\newcommand{\desco}{D\hspace{-4mm}\not\hspace{2mm}}

%%%%

%\draft command makes pacs numbers print
%\draft

\title{\boldmath ALPs and Heavy Hadron Chiral Perturbation Theory }

\author{Namit Mahajan
}
\email{nmahajan@prl.res.in}
\affiliation{
 Theoretical Physics Division, Physical Research Laboratory, Navrangpura, Ahmedabad
380 009, India
}

%\date{\today}

\begin{abstract}
Axion Like Particles (ALPs) belong to a well motivated class of particles which are part
of a more complete UV theory. In this work, we initiate and present the inclusion of such ALPs or the conventional axions in the Heavy Hadron Chiral Perturbation Theory (HHChiPT) in a model independent effective lagrangian way. Such a framework is well suited for studying heavy to heavy quark flavour preserving transitions. For light ALPs, the limits on the ALP couplings (or their combinations) turn out to be independent of ALP mass. Radiative and leptonic decays of heavy vector mesons to corresponding heavy pseudoscalar mesons will be able to provide some of the most stringent constraints on ALPs, particularly the diagonal couplings.
\end{abstract}

%\pacs{
%}
\maketitle
%\narrowtext

%\section{Introduction}
The Standard Model (SM) of particle physics allows the $ {\theta}$-term: $\theta G\tilde{G}$, where $G$ denotes the gluon field strength. Such a term, then has important phenomenological implications. The neutron electric dipole moment constrains $\bar{\theta} \leq 10^{-10}$, where $\bar{\theta}$ is the physical parameter related to $\theta$ via chiral rotation of the quark fields. This is commonly called the strong CP problem (see \cite{Dine:2000cj}-\cite{Baluni:1978rf} for more details) since $\bar{\theta}\to 0$ does not enhance  the symmetry of the theory. One thus expects $\bar{\theta} \sim \mathcal{O}(1)$. Axions provide an elegant solution to the strong CP problem \cite{Peccei:1977hh}-\cite{Wilczek:1977pj}. See also \cite{Peccei:2006as}, \cite{Kim:2008hd}. They are pseudo-Goldstone particles arising due to the breaking of a global symmetry, the Peccei-Quinn (PQ) symmetry. In the original QCD axion set up, they acquire a small mass after QCD confines, and thus the axion mass and decay constant get related to the pion mass and decay constant: $m_a f_a = m_{\pi}f_{\pi}$ (see \cite{Kim:1979if}-\cite{Zhitnitsky:1980tq} for concrete realizations). While this is rather interesting, it was also realised \cite{Rubakov:1997vp} that other solutions to the strong CP problem are achievable with heavier axions or axion like particles (ALPs). Since we are interested in a model independent framework, the term ALP is used interchangably for the axion below.  In more generic models, ALPs may or may not end up solving the strong CP problem. Often, the term axion or ALP is employed to indicate a particle like the original QCD axion i.e. a particle with couplings like the QCD axion but not necessarily providing a (complete) solution to the strong CP problem. A rough way to see this could be by recalling that there are no global symmetries in a theory of gravity (see \cite{Coleman:1989zu}-\cite{Kallosh:1995hi}). This means that at low energies there would be suppressed operators such that the global symmetry is an accidental approximate symmetry of the theory (\cite{Holman:1992us}-\cite{Ghigna:1992iv}). In such a case, these operators would introduce extra correction terms to the axion/ALP potential, which, in the absence of these corrections, would have provided a solution to the strong CP problem, thus requiring these operators to be highly suppressed for the solution to approximately/practically hold. With this rough reasoning in mind, we proceed without questioning the exact extent to which the ALPs ameliorate the strong CP problem. There have been extensive studies on ALPs and their phenomenlogical signatures. The mass of the ALPs can range from sub-MeV to several GeVs. The astrophysical observations typically provide very strict constraints on the sub-MeV or very light ALPs whie collider searches are well suited for multi-GeV ALP masses (for an incomplete list of constraints/bounds on axion/ALP couplings, see \cite{Preskill:1982cy}-\cite{Carenza:2020cis}). 

As the original motivation suggests, ALPs have couplings with the gluons. In specific realizations of the ALP/axion models, the SM fermions may be uncharged under the PQ symmetry (KSVZ models) while in the DSFZ models there are additinal couplings to fermions. Similar additional ALP couplings to fermions are also present in the supersymmetric or composite Higgs type models, see \cite{Bellazzini:2017neg}, \cite{Gripaios:2009pe}. In a model independent set up, a complete set of operators with couplings suppressed by the typical heavy scale is considered without any direct contact to a specific model. ALPs would generally also have flavour dependent, and possibly flavour violating, couplings to the fermions \cite{Davidson:1981zd}-\cite{Calibbi:2016hwq} either due to the choice of PQ charges or via radiative mechanisms. This then opens up the possibility to study and constrain ALPs via flavour observables. See \cite{Bauer:2021mvw} for a very detailed discussion of ALP probes via flavour observables. For an overview of axion models and generalities, see \cite{DiLuzio:2020wdo}.

In a UV complete model, there are additional fields beyond those in the SM. The model assigns specific quantum numbers or charges to different fields, SM or beyond, and a consistent set of interaction terms is written. From the weak scale physics perspective, the heavier degrees of freedom are systematically integrated out and following the standard procedure of renormalization group evolution equations and running (RGEs), the effective Wilson coefficients are determined at the weak scale (see \cite{Chala:2020wvs} for RG running for ALPs). Alternatively, we could start just above the weak scale and add to the SM lagrangian a set of terms involving SM fields and only the axion, with no explicit mention of any additional fields. The ALP (or axion) couples to the Higgs and chiral fermion fields via the derivative couplings to a set of gauge invariant operators, while the couplings to the gauge bosons have the form of the $\theta$-term. Proceeding this way may be advantageous since the original fields before the PQ symmetry breaking would mix in appropriate linear combinations after the PQ symmetry is broken. Unless there is a concrete UV complete model to work with, this may be a better approach particularly when the interest is in ALPs and not the other particles of the extended model. The most general set of interaction terms at the weak scale then reads \cite{Georgi:1986df}-\cite{Bauer:2021wjo}:
\begin{eqnarray}
\mathcal{L}_{int} &=& \frac{\partial^{\mu}a}{F}\left(c_{\phi}\phi^{\dag}iD_{\mu}\phi + \sum_f c_f \bar{f}\gamma_{\mu}f\right) \nonumber\\
&+& \sum_V c_{VV}\frac{\alpha_V}{4\pi}\frac{a}{F}V_{\mu\nu}\tilde{V}^{\mu\nu}\, ,
\end{eqnarray}
where the sum in the first line runs over all the chiral fermions, $f$, while $V$ in the second line runs over gluons ($A_{\mu}^a$), $SU(2)_L$ and $U(1)_Y$ gauge bosons, $W_{\mu}^i$ and $B_{\mu}$ respectively. $4\pi F = \Lambda_{PQ}$ is the PQ symmetry breaking scale. The coupling to the Higgs boson is actually a redundant operator and can be removed in favour of the fermion-ALP interaction term, plus a term which vanishes by equation of motion. Further, below the weak scale, after the electroweak symmetry breaking, $W$ and $Z$ are not the dynamical degrees of freedom. Thus, the effective theory at scales below the weak scale only has ALP couplings to light fermions, gluons and photons. The ALP couples to photons with a strength, $c_{\gamma\gamma}$, which is a linear combination of $c_{WW}$ and $c_{BB}$. The quantities $c_f$ are Hermitian matrices: different for the left and right chiral fermions. Possible off-diagonal entries in these matrices would lead to flavour changing transitions. At the mesonic level, the above lagrangian is matched on to an effective chiral lagrangian by first removing the ALP-gluon interaction term, accomplished through a chiral rotation of the quark fields:
\begin{equation}
 q(x) \rightarrow \mathrm{exp}\left[-ic_{GG}(\boldsymbol{\delta}_q + \boldsymbol{\kappa}_q)\frac{a(x)}{F}\right]q(x)\, ,
\end{equation}
where the Hermitian matrices $\boldsymbol{\delta}_q$ and $\boldsymbol{\kappa}_q$ are chosen to be diagonal in the mass basis. Imposing $Tr(\boldsymbol{\kappa}_q) = 1$ completely eliminates the ALP coupling to gluons while at the same time modifying couplings to photons and quarks. The pseudoscalar mesons are contained in the field $\boldsymbol{\Sigma}$: 
\begin{equation}
 \boldsymbol{\Sigma}(x) = \mathrm{exp}\left[2i\,\boldsymbol{M}/f_{\pi}\right]\, ,
\end{equation}
where $\boldsymbol{M} = T^a \Pi^a$ is the matrix containing the octet mesons, with $T^a$ being the Gell-Mann matrices, and $\Pi^a$ are the pseudoscalar mesons. The covariant derivative acting on $\boldsymbol{\Sigma}$ allows to include the derivative couplings of the ALPs with the chiral quark fields. Denoting the original left and right handed derivative quark couplings by $\boldsymbol{k}_Q$ and $\boldsymbol{k}_q$, the modified couplings after the above chiral rotation on the quark fields is performed are given by:
\begin{equation}
 \hat{\boldsymbol{k}}_{Q(q)} = \mathrm{exp}(i\varphi^{-(+)}_{q}a/F)[\boldsymbol{k}_{Q(q)} + \varphi^{-(+)}_{q}]\mathrm{exp}(i\varphi^{-(+)}_{q}a/F)\, ,
\end{equation}
where 
\begin{equation}
 \varphi^{\pm} = c_{GG}\,(\boldsymbol{\delta}_q \pm \boldsymbol{\kappa}_q).
\end{equation}
The covariant derivative on the field $\boldsymbol{\Sigma}$ then reads
\begin{equation}
 i\boldsymbol{D}_{\mu}\boldsymbol{\Sigma} = i\partial_{\mu}\boldsymbol{\Sigma} + e\,A_{\mu}[\boldsymbol{Q},\boldsymbol{\Sigma}] + \frac{\partial_{\mu}a}{F}(\hat{\boldsymbol{k}}_{Q}\,\boldsymbol{\Sigma} - \boldsymbol{\Sigma}\,\hat{\boldsymbol{k}}_{q})\, ,
\end{equation}
with $\boldsymbol{Q} =$ Diagonal$\,(Q_u,\,Q_d,\,Q_s)$ are the fractional quark charges. It was pointed out in \cite{Bauer:2021wjo} that the derivative coupling above, which was missed in the original work \cite{Georgi:1986df} and in subsequent studies, has important consequences for $K\to\pi a$ modes. Below, we'll find the same derivative coupling at work. Further, if one assumes Minimal Flavour Violation \cite{DAmbrosio:2002vsn}, the diagonal quark couplings respect the relations \cite{Bauer:2020jbp}:
\begin{equation}
 c_{ss} = c_{dd}, \,\,\,\,\,\, ({\boldsymbol{k}}_d)_{11} - ({\boldsymbol{k}}_d)_{22} = 0 = ({\boldsymbol{k}}_D)_{11} - ({\boldsymbol{k}}_D)_{22}.
\end{equation}
The flavour observables usually considered probe the off-diagonal couplings. Which processes or observables can probe the diagonal couplings? To this end, we consider heavy vector mesons ($D^*,\, B^*$) decaying to heavy pseudoscalar mesons ($D,\, B$) and an ALP. For the heavy to heavy transitions involving additional light mesons, the appropriate framework is the Heavy Hadron Chiral Perturbation Theory (HHChiPT) \cite{Wise:1992hn}-\cite{Cho:1992nt}. This is what we shall follow below since the ALP coupling is automatically included via the covariant derivative defined above. The non-derivative couplings of ALPs to the quarks and mesons gets included via the ALP field dependent quark mass matrix:
\begin{equation}
 \hat{\boldsymbol{m}}_q(a) = \mathrm{exp}\left(-2i{\boldsymbol{k}}_q\,c_{GG}\,\frac{a}{F}\right)\boldsymbol{m}_q\, ,
\end{equation}
where $\boldsymbol{m}_q = $ Diagonal$\,(m_u,\,m_d,\,m_s)$. The physical ALP field will mix with the neutral pseudoscalars. At the leading order in $1/F$, the ALP-$\pi^0$ mixing angle after making a judicious choice, ${\boldsymbol{k}}_q = \boldsymbol{m}_q^{-1}/Tr[\boldsymbol{m}_q^{-1}]$, turns out to be proportional to ALP mass squared: $\theta_{a\pi} \propto \frac{f_{\pi} m_a^2}{F(m_{\pi}^2-m_a^2)}\,(\hat{c}_{uu}-\hat{c}_{dd})$, where $\hat{c}_{qq} = k_q - k_Q + 2 \kappa_q\,c_{GG}$. The combination $\Delta\hat{c} = \hat{c}_{uu}-\hat{c}_{dd}$ is isospin violating and therefore, $\theta_{a\pi}$ is an isospin violating factor. A rough estimate of this factor could be obtained from the isospin violating factor $\frac{m_d - m_u}{m_s - (m_u+m_d)/2} \simeq 0.02$ which enters the isospin violating decay $D_s^* \to D_s\pi^0$ via the $\eta-\pi$ mixing \cite{Cho:1994zu}. Thus we naively have $\Delta\hat{c} \sim \mathcal{O}(0.02)$. For light ALPs, $m_a << m_{\pi}$, this mixing is rather insignificant. The situation is different for $m_a$ close to $m_{\pi}$.

Chiral Perturbation Theory (ChiPT) is well suited for describing the physics of the pseudo-Goldstone bosons forming the octet mesons (see \cite{Gasser:1983yg}-\cite{Scherer:2002tk}). The other extreme limit is that of hadrons containing a heavy quark of mass $m_Q$. In the infinite mass limit, $m_Q\to \infty$, hadrons containing a heavy quark form degenerate doublets of total spin. As an example, and what is relevant for the present purpose, for the light degrees of freedom having spin $1/2$, there will be spin zero and spin one states which are degenerate in mass. Loosely speaking, in the infinite mass limit, the extend of the heavy quark (Compton wavelength $\sim 1/m_Q$) is much smaller than the resolving power of the light degrees of freedom ($\sim 1/\Lambda_{QCD}$), and therefore the spin orientation of the light degrees of freedom plays no role in determining the gross properties. This then allows for a systematic expansion in $1/m_Q$ yielding corrections to this extreme limit which can be reliably calculated. The effectiveness is clearly seen from the following mass differences \cite{ParticleDataGroup:2022pth}:
\begin{equation}
 m_{B^*} - m_B \sim 45\, {\mathrm MeV}, \,\,\,\,m_{D^*} - m_D \sim 142\, {\mathrm MeV} > m_{\pi}
\end{equation}
Very evidently, the heavy quark symmetry is expected to work much better in the bottom quark sector than the charm quark. Though $m_c > \Lambda_{QCD}$ and thus reliable results based on heavy quark symmetry (and accounting for systematic corrections in $1/m_c$) can be obtained, the fact that $m_{D^*} - m_D > m_{\pi}$, while the same is not true for B-mesons has important implications: $D^* \to D\pi$ decays are allowed while no such decays are kinematically possible for $B^*$. However, $D^*$ ($B^*$) decaying to $D(B)$ and an ALP is possible. For mesons having a heavy quark $Q$ and light (anti-)quark $\bar{q}_a$ ($a=u,\,d,\,s$), the pseudoscalar mesons are denoted by $P_a$ while the vector mesons are denoted by $P_a^*$. We have the multiplets: $(D^0,\, D^+,\, D_s)$, $(D^{*0},\, D^{*+},\, D^*_s)$ and $(B^-,\, B^0,\, B_s)$, $(B^{*-},\, B^{*0},\, B^*_s)$. The velocity of the heavy fields is denoted by $v_{\mu}$. It is convenient to combine these mesons in a $4\times 4$ matrix, $\boldsymbol{H}_a$:
\begin{equation}
 \boldsymbol{H}_a = \left(\frac{\not{v}+1}{2}\right)[-P_a\gamma_5 + P_a^{*\mu}\gamma_{\mu}].
\end{equation}
Employing $\boldsymbol{H}_a$ and $\boldsymbol{\Sigma}$, the most general $SU(3)_L\times SU(3)_R$ invariant terms can be written.
\begin{eqnarray}
 \mathcal{L}_{\mathrm {HHChiPT}} &=& -i\,Tr[\bar{\boldsymbol{H}}_av_{\mu}\partial^{\mu}\boldsymbol{H}_a] \nonumber \\
 &+& \frac{i}{2}Tr[\bar{\boldsymbol{H}}_a\boldsymbol{H}_bv_{\mu}(\boldsymbol{\Sigma}^{\dag}\partial^{\mu}\boldsymbol{\Sigma} + \boldsymbol{\Sigma}\partial^{\mu}\boldsymbol{\Sigma}^{\dag})_{ba}] \\
 &+& \frac{i}{2}g\,Tr[\bar{\boldsymbol{H}}_a\boldsymbol{H}_b\gamma_{\mu}\gamma_5(\boldsymbol{\Sigma}^{\dag}\partial^{\mu}\boldsymbol{\Sigma} - \boldsymbol{\Sigma}\partial^{\mu}\boldsymbol{\Sigma}^{\dag})_{ba}] + .... \nonumber
\end{eqnarray}
It is then straightforward to expand in powers of $\boldsymbol{M}$ to obtain the Feynman rules for light pseudoscalars $\pi,\,K, \,\eta$ interacting with heavy mesons. The coupling strength $g$ is independent of the heavy quark mass $m_Q$ owing to the heavy quark flavour symmetry. What this means is that when employing the HHChiPT lagrangian, its the same coupling $g$ that will enter the $D^*$ and $B^*$ decays. It is instructive to remind that in the heavy mass limit, the coupling $g$ (also denoted by $\hat{g}$ in the literature) is related to  $g_{P^*P\pi}$ coupling (computed by means of the $P^*P\pi$ correlation function) via the relation $\lim_{m_Q\to\infty} g_{P^*P\pi}/(2 m_P) = g/f_{\pi}$. Estimates for $g$ range from $0.3$ to $0.49$ and are sensitive to the method used to compute (see for example \cite{Khodjamirian:2020mlb}). This relation is expected to get corrections as strictly speaking $m_c \neq m_b$. To linear order in $\boldsymbol{M}$, $\mathcal{L}_{\mathrm HHChiPT}$ yields
\begin{eqnarray}
 \mathcal{L}_{\mathrm {int}} &=& -\frac{g}{f_{\pi}}\,Tr[\bar{\boldsymbol{H}}_a\boldsymbol{H}_b\gamma_{\mu}\gamma_5\partial^{\mu}\boldsymbol{M}_{ba}] \nonumber \\
 &=& \left(-\frac{2g}{f_{\pi}}\partial^{\mu}\boldsymbol{M}_{ba}P_a^{\dag}P_{b\mu}^* + h.c.\right) \\
 &+& \left(\frac{2ig}{f_{\pi}}\right)\partial^{\mu}\boldsymbol{M}_{ba}P^{*\alpha\dag}_bP_{a}^{*\beta}v^{\lambda}\epsilon_{\alpha\lambda\beta\mu} + .... \nonumber
\end{eqnarray}
where ellipses denote terms higher than quadratic in $P,\,P^*$ which are not needed for the present purpose. It is to be noted that there are no $P_aP_bM_{ab}$ terms due to parity argument. This interaction lagrangian then immediately results in $P^* \to P \pi$ decay rate (applicable only for $D^*$ decay):
\begin{equation}
 \Gamma(P^* \to P\pi^a) = \frac{g^2}{6\pi f_{\pi}^2}\delta^a\,\vert\vec{p}_{\pi}\vert^3\, ,
\end{equation}
with $\delta^a = 1/2$ for $\pi^0$ in the final state and unity otherwise, and $\vert\vec{p}_{\pi}\vert = \frac{1}{2m_{P^*}}\left([m_{P^*}^2-(m_P+m_{\pi})^2)][m_{P^*}^2+(m_P+m_{\pi})^2)]\right)^{1/2}$. 

It is now straightforward to include ALPs in the interactions by making the replacement $i\partial^{\mu} \rightarrow iD^{\mu}$ in $\mathcal{L}_{\mathrm {HHChiPT}}$, with the covariant derivative define above. After some algebra and making use of the hermiticity of $\hat{\boldsymbol{k}}_{Q(q)}$, the ALP interaction term reads
\begin{equation}
 \mathcal{L}_{\mathrm {ALP}} = g\frac{\partial^{\mu}a}{F}\,Tr[\bar{\boldsymbol{H}}_a\,\boldsymbol{H}_b\gamma_{\mu}\gamma_5\,(\hat{\boldsymbol{k}}_{Q} - \hat{\boldsymbol{k}}_{q})_{ba}].
\end{equation}

Comparing $\mathcal{L}_{\mathrm {int}}$ and $\mathcal{L}_{\mathrm {ALP}}$, it is easy to see that the same HHChiPT coupling $g$ enters both, and further the structure of the interaction terms is very similar. $\mathcal{L}_{\mathrm {ALP}}$ provides a direct derivative term for $P^*_a\to P_a +\,{\mathrm {ALP}}$ while the ALP-$\pi^0$ mixing provides an indirect channel for the ALP production. Because of charge conservation, it is the combination of the diagonal couplings that enters $\Gamma(P^*_a\to P_a + \,{\mathrm {ALP}})$ i.e. $a=b$. 

\begin{figure}[ht!]
\vskip 0.32cm
\hskip 1.35cm
\hbox{\hspace{0.03cm}
\hbox{\includegraphics[scale=0.5]{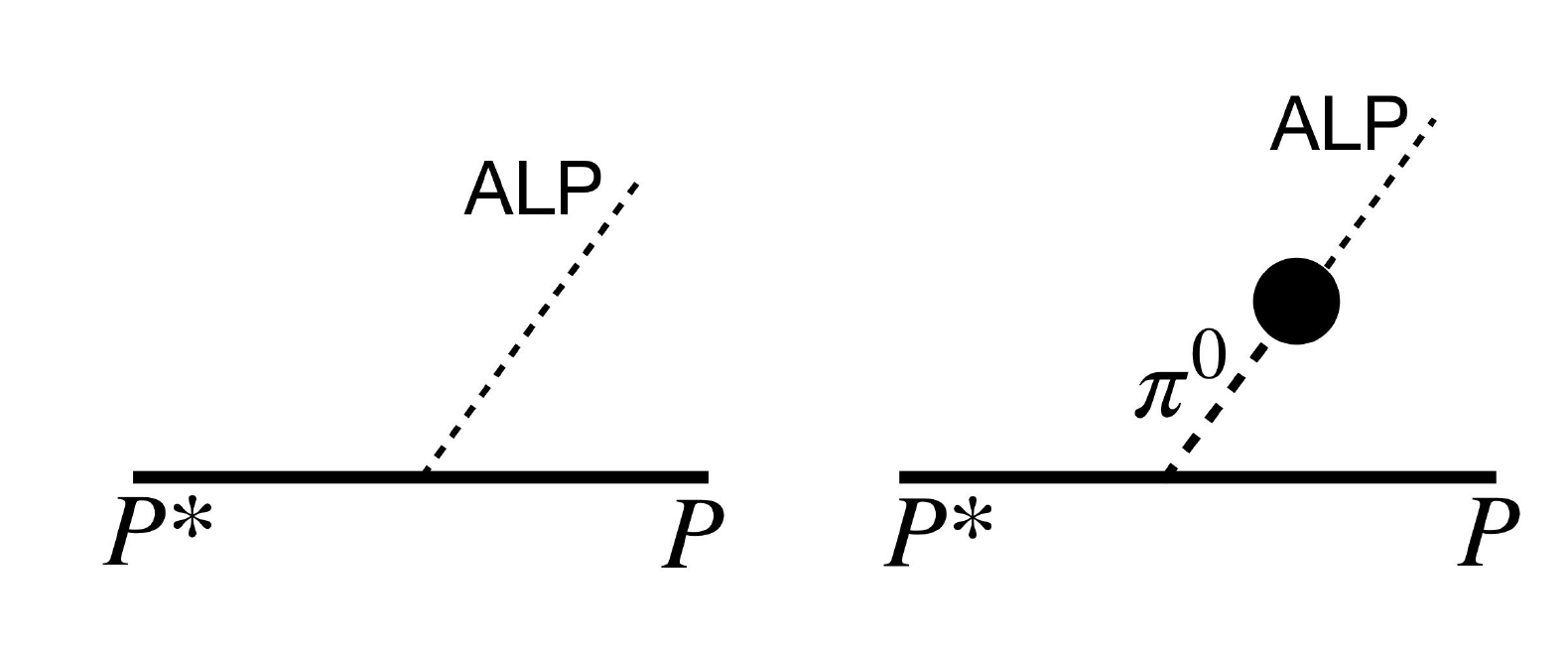}}
}
\caption{Feynman diagrams for a heavy vector decaying into a heavy pseudoscalar meson and ALP. Left: direct ALP coupling. Right: ALP-$\pi^0$ mixing. The blob denotes $\theta_{a\pi}$.}
\label{fig1}
\end{figure}

For $m_a<< m_{\pi}$, when the ALP-$\pi^0$ mixing is negligible, the decay width for $P^*_a\to P_a +\,{\mathrm ALP}$ is dominated by the direct ALP derivative term in the interaction lagrangian, and is given by
\begin{eqnarray}
 \Gamma(P^*_a\to P_a +\,{\mathrm {ALP}}) &=& \frac{g^2}{6\pi F^2}\,\vert(\hat{\boldsymbol{k}}_{Q} - \hat{\boldsymbol{k}}_{q})_{aa}\vert^2\,\vert\vec{p}_{a}\vert^3 \nonumber \\
 &=& 2 \left[\frac{f_{\pi}^2 \vert\vec{p}_{a}\vert^3}{F^2 \vert\vec{p}_{\pi}\vert^3}\right]\,\vert(\hat{\boldsymbol{k}}_{Q} - \hat{\boldsymbol{k}}_{q})_{aa}\vert^2\nonumber \\&\otimes& \Gamma(P^*_a\to P_a\pi^0)\, ,
\end{eqnarray}
where the second line holds only for the charm mesons. For very light ALPs, $\vert\vec{p}_{a}\vert = \frac{(m_{P^*}^2-m_P^2)}{2m_{P^*}}$. Thus, $\vert\vec{p}_{a}\vert \sim 145 \,\mathrm{MeV}$ for $D^*$ decays while it is $\sim 45 \,\mathrm{MeV}$ for $B^*$ decays. Using the above, we therefore have the following relations:
\begin{equation}
 \Gamma(B_a^*\to B_a +\,{\mathrm {ALP}}) = \underbrace{\left(\frac{\vert\vec{p}_{a}\vert_{B^*}}{\vert\vec{p}_{a}\vert_{D^*}}\right)^3}_{\sim 0.03}\,\Gamma(D_a^*\to D_a +\,{\mathrm {ALP}}).
\end{equation}
The absolute decay width is known only for $D^{*+}$ \cite{ParticleDataGroup:2022pth}: $\Gamma^{D^{*+}}_{\mathrm tot} = 83.4\pm1.8\,\, {\mathrm {keV}}$, while for $D^{0*}$ and $D_s^{*}$ only upper limits are available. To get some idea, we make use of $\Gamma^{D^{*+}}_{\mathrm tot}$ and individual branching ratios:
\begin{eqnarray}
 BR(D^{*+}\to D^0 \pi^+) &=& (67.7\pm 0.5)\%, \nonumber \\
 BR(D^{*+}\to D^+ \pi^0) &=& (30.7\pm 0.5)\%, \\
 BR(D^{*+}\to D^+ \gamma) &=& (1.6\pm 0.4)\%. \nonumber
\end{eqnarray}
These very simply add to $100\%$ with an error of about $0.8\%$. For the sake of illustration, assuming a difference of $0.5\%$ between the total and sum of the individual channels, provides an upper limit on $\Gamma(D^{*+}\to D^+ +\,{\mathrm ALP}) < 0.4 \,\,{\mathrm {keV}}$. This in turn yields, $\Gamma(B^{*0}\to B^0 +\,{\mathrm ALP}) < 0.012\,\, {\mathrm {keV}}$. Isospin or flavour $SU(3)$ symmetry can further be used to obtain an estimate of other decay widths. Using the above estimate, in conjunction with the second line of Eq.(15), one obtains a constraint on the ALP couplings by assuming a value for the ALP decay constant $F$. For the QCD axion, $1/F = -1/(2 c_{GG} F_a)$. For $m_a<< m_{\pi}$, these limits and constraints are essentially independent of the ALP mass. 

We now consider the impact due to $\theta_{a\pi}$ i.e. the diagram on the right in Fig. (1). The rate due to ALP-pion mixing then reads
\begin{eqnarray}
 \Gamma(P^*_a\to P_a +\, \mathrm{ALP}) &=& \frac{g^2 \theta_{a\pi}^2}{12\pi f_{\pi}^2}\vert\vec{p}_{a}\vert^3  \nonumber \\
 %&=& \frac{g^2 \theta_{a\pi}^2}{6\pi F^2} \left[\frac{m_a^2 \vert(\hat{c}_{uu}-\hat{c}_{dd})\vert^2}{16(m_{\pi}^2-m_a^2})\right] \nonumber \\
 &\leq& 10^{-5}\,\frac{g^2}{6\pi F^2}\left[\frac{m_a^2}{m_{\pi}^2-m_a^2}\right]\, .
\end{eqnarray}
where use has been made of the crude estimate of the isospin violating factor $\Delta\hat{c} = \hat{c}_{uu}-\hat{c}_{dd}$. The huge suppression can only be compensated for $m_a$ very close to $m_{\pi}$. Away from this, the suppression factor makes the ALP-pion mixing contribution completely insignificant.

For the same reason that there are no $P_aP_BM_{ab}$ terms, there are no terms with the light pseudoscalar mesons replaced with the ALP field. However, the ALP included HHChiPT lagrangian yields terms quadratic in heavy vector mesons and ALP. Since the mass difference between heavy vector mesons containing s-quark and d-quark both in the charm ($\sim 100 \,\mathrm{MeV}$) and bottom sector ($\sim 90 \,\mathrm{MeV}$) is smaller than the pion mass, $P^*_s\to P^*_d +\, \mathrm{ALP}$ is possible while pionic modes are kinematically forbidden. The amplitude for such a process reads
\begin{equation}
 A(P^*_s\to P^*_d +\, \mathrm{ALP}) = \frac{2g}{F}(\hat{k}_Q-\hat{k}_q)_{23}\,p_a^{\mu}\varepsilon_2^{\alpha}\varepsilon_3^{\beta}v^{\lambda}\epsilon_{\alpha\lambda\beta\mu}\, .
\end{equation}
The combination of ALP-quark couplings entering the above set of amplitudes is the same that enters $B_s\to \mu\mu$ amplitude along with ALP coupling to muons. 

\begin{figure}[ht!]
\vskip 0.32cm
\hskip 1.35cm
\hbox{\hspace{0.03cm}
\hbox{\includegraphics[scale=0.5]{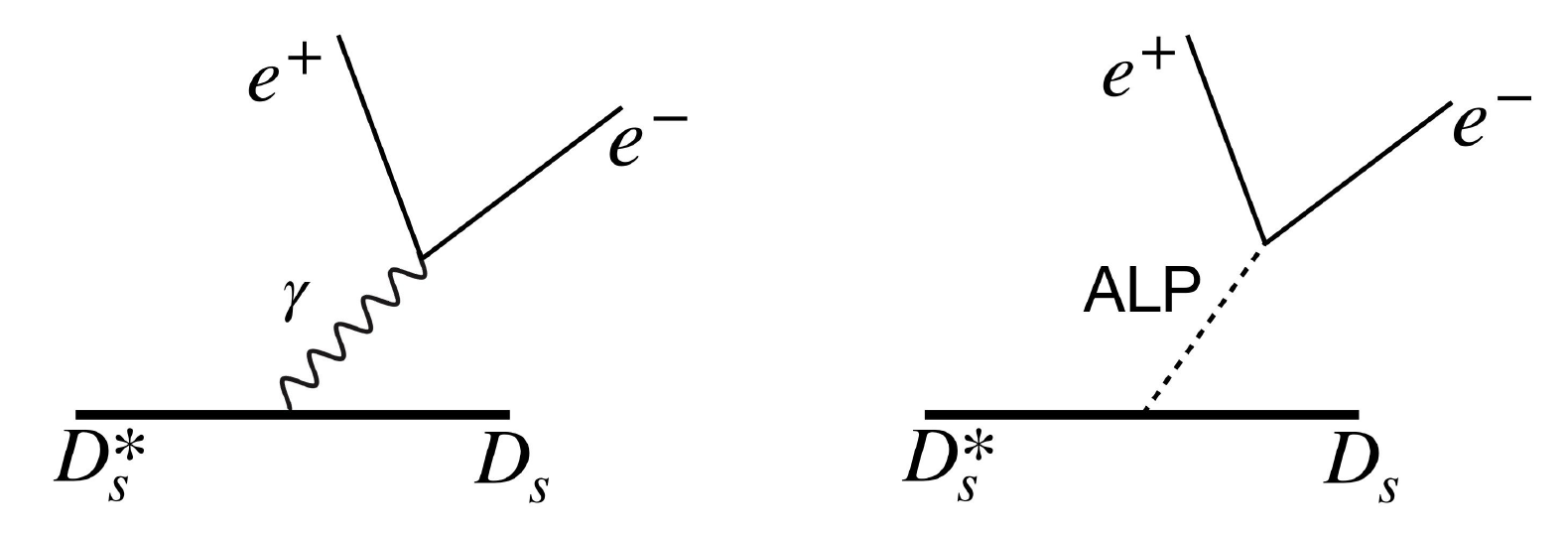}}
}
\caption{Feynman diagrams for a heavy vector decaying into a heavy pseudoscalar meson and ALP. Left: direct ALP coupling. Right: ALP-$\pi^0$ mixing.The blob denotes $\theta_{a\pi}$.}
\label{fig2}
\end{figure}

We next turn to $D_s^*\to D_s\,e^+e^-$. The branching ratio for the decay mode is $(6.7\pm 1.6)\times 10^{-3}$ and the branching ratio for the radiative mode is $(93.5 \pm 0.7)\%$ \cite{ParticleDataGroup:2022pth}. Decay to a pion in the final state is isospin violating and therefore the radiative decay dominates. A rough estimate for the branching ratio can be made by multiplying the branching ratio for the radiative $D_s^*$ decay by $\alpha_{em}$. We thus obtain: $0.935\,\alpha_{em} \sim 6.8\times 10^{-3}$ which is rather in good agreement with the leptonic branching ratio quoted above. This then provides a good avenue to constrain the ALP quark and electron couplings. Alternatively, using information on the ALP-electron coupling from other mesonic decays and lepton flavour violating processes, one can obtain constraints on $(\hat{k}_Q-\hat{k}_q)_{33}/F$. The leptonic channel is also expected to play an important role in case of $B_s^* \to B_s$ transitions since pionic final state is kinematically forbidden. The radiative decay is expected to saturate the width but there would be some window for the leptonic channel. A precise measurement of the readiative and leptonic modes will help set stringent bounds on the ALP couplings.

The next step would be to consider ALP interactions at the loop level. This would include not just renormalizing the HHChiPT lagrangian at the one loop level but also ALP contribution to $P^*\to P\gamma$ and $P^*\to P\pi$ processes via loops. Some representative diagrams are shown in Fig.(3). A proper study at the one loop level would require a careful expansion of the HHChiPT and ChiPT lagrangians in order to include the relevant ALP interactions, including possible vertices with ALP and pion or photon together with heavy mesons.

\begin{figure}[ht!]
\vskip 0.32cm
\hskip 1.35cm
\hbox{\hspace{0.03cm}
\hbox{\includegraphics[scale=0.35]{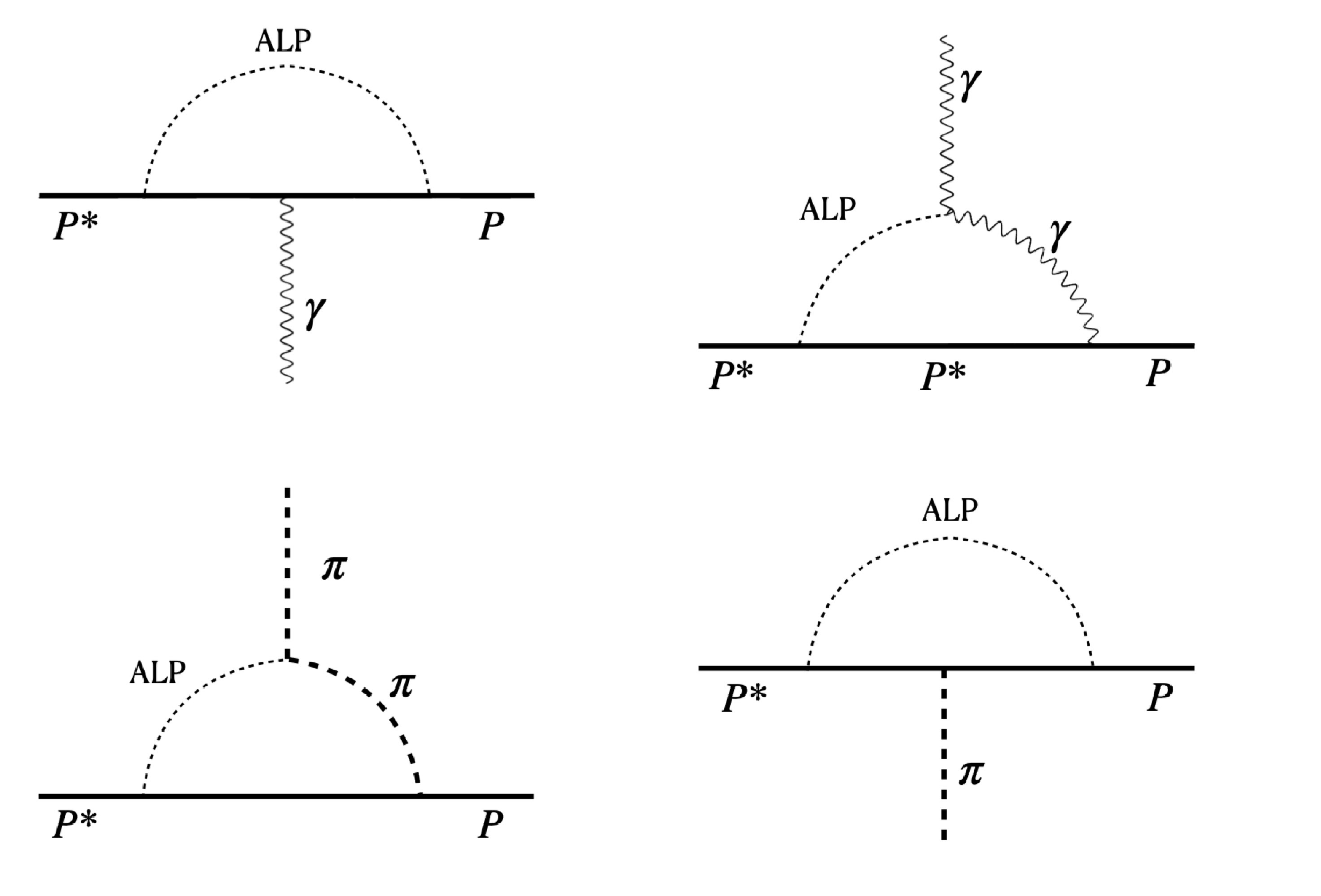}}
}
\caption{Feynman diagrams for a heavy vector decaying into a heavy pseudoscalar meson and ALP. Left: direct ALP coupling. Right: ALP-$\pi^0$ mixing.The blob denotes $\theta_{a\pi}$.}
\label{fig3}
\end{figure}

In the present work, we've initiated a study of ALPs interacting with heavy mesons by systematically incorporating the ALPs in the HHChiPT. Utilising the leading interactions, and the available branching ratios for the heavy vector meson decays, decay rates to final states containing an ALP are estimated. It is shown that the constraints are essentially independent of ALP mass if $m_a<< m_{\pi}$ and the ALP-pion mixing plays no role Only when $m_a \to m_{\pi}$, there is a contribution from the mixing term. For heavier ALPs, the ALP mass would enter the decay rate through $\vert\vec{p}_a\vert$ which will be defined similar to $\vert\vec{p}_{\pi}\vert$. We've restricted ourselves to a general model independent framework for the ALP interactions. In a given ALP or axion model, the specific non-zero couplings of the ALPs with the SM fields can then be considered and RG evolved to the lower scales. The RG evolution would also generate more couplings at the lower scale even if they were zero to start with. In a concrete UV model, due to the structure of the model, there may be one or two non-zero couplings at the high scale, which would generate other couplings at the lower scales, and direct constraints could be put on those couplings. A detailed study on specific ALP/axion models, including the one loop contributions properly taken into account via the ALP-HHChiPT and ALP-ChiPT interactions is left for a future study. 

%The focus will be on the $\rho$ exchange diagrams, namely diagram (b) and (c). 

%expt prospect
%\vskip 3cm

%MFV

%

\end{document}